\documentclass[twocolumn]{aastex631}
\def\xmm{{\it XMM-Newton }}
\def\cha{{\it Chandra }}

\def\nus{{\it NuSTAR }}

\shorttitle{Long-term spin-down trend of ULXP M82 X-2}
\shortauthors{Liu J.}

\begin{document}
\title{The long-term spin-down trend of ultra-luminous X-ray pulsar M82 X-2}
\correspondingauthor{Jiren Liu}
\email{jrliu@swjtu.edu.cn}
\author{Jiren Liu}
\affiliation{Beijing Planetarium, Beijing Academy of Science and Technology, Beijing 100044, China}
\affiliation{School of Physical Science and Technology, Southwest Jiaotong University, Chengdu Sichuan 611756, China}



\begin{abstract}

The discovery in 2014 of the pulsation from the ultra-luminous X-ray source (ULX) M82 X-2
in 2014 has changed our view of ULXs. 
Because of the relatively short baseline over which pulsations have been detected so far,
M82 X-2's spin state had been assumed to be in an equilibrium state.
Using \cha and \xmm archive data, we are able to investigate the pulsation 
of M82 X-2 back to 2005 and 2001. The newly determined spin frequencies clearly show 
a long-term spin-down trend.
If this trend is caused by magnetic threading, we infer a 
dipolar magnetic field of $\sim1.2\times10^{13}$ G and that a mild beaming factor 
($\sim4$) is needed to match the braking torque with the mass accretion torque. 
On the other hand, there are \nus observations showing instantaneous spin-down behaviours, 
which might favour a variable prograde/retrograde flow scenario for M82 X-2.

\end{abstract}


\section{Introduction}

The discovery of X-ray pulsation from the ultra-luminous X-ray source (ULX) M82 X-2
revealed the existence of accreting 
magnetized neutron star with an apparent luminosity of around $10^{40}$erg\,s$^{-1}$ \citep{Bac14}.
Since then, a few more ultra-luminous X-ray pulsars (ULXPs) have been identified,
e.g., NGC 5907 X-1 \citep{Isr17a}, NGC 7793 P13
\citep[][]{Fur16,Isr17b}, NGC 300 X-1 \citep{Car18}, and NGC 1313 X-2 \citep{Sat19}.
It is plausible that neutron stars, instead of black holes, are the dominant accretors 
of the ULX population \citep[for a recent review, see][]{Kin23}.

Despite being the first discovered ULXP, many properties of M82 X-2 remain unclear.
One key property, concerning how the observed luminosity is related to the intrinsic luminosity,
is still debated. \citet{Mus21} showed that a large pulsed fraction ($>10\%$) 
implies no strong beaming of ULXPs. Recently, \citet{Bac22} determined an orbital 
decay rate of $-5.69\times10^{-8}$\,s\,s$^{-1}$ for M82 X-2. This implies a mass transfer 
rate of $\sim200$ times the Eddington limit, which is sufficiently high to produce the 
observed luminosity and requires no strong beaming.
Another unsolved problem concerns the magnetic field of M82 X-2.
To reach a luminosity as high as $10^{40}$erg\,s$^{-1}$, a strong surface 
magnetic field ($\sim10^{14}$\,G) is needed to reduce the electron 
scattering opacity.
On the other hand, many studies had assumed spin equilibrium of M82 X-2 to estimate its
dipolar magnetic field \citep[$\sim 10^{13}-10^{14}$ G, e.g.][]{Bac22}.

In this paper, we report the detection of pulsation from M82 X-2 back to 2005 (\cha data), 
and 2001 (\xmm data). 
The subarray mode of \cha ACIS-S has a time resolution 
of 0.44104\,s, which makes the detection of a pulse period around 1.35\,s possible.
The \xmm EPIC-pn data has a full frame time resolution of 0.07 s, which is also 
high enough for the detection of pulsation from M82 X-2.
The newly measured spin frequencies of M82 X-2 show a surprising long-term spin-down 
trend, and enable an estimation of the dipolar magnetic field of M82 X-2
based on the braking torque.

\begin{figure*}
	\includegraphics[width=3.2in]{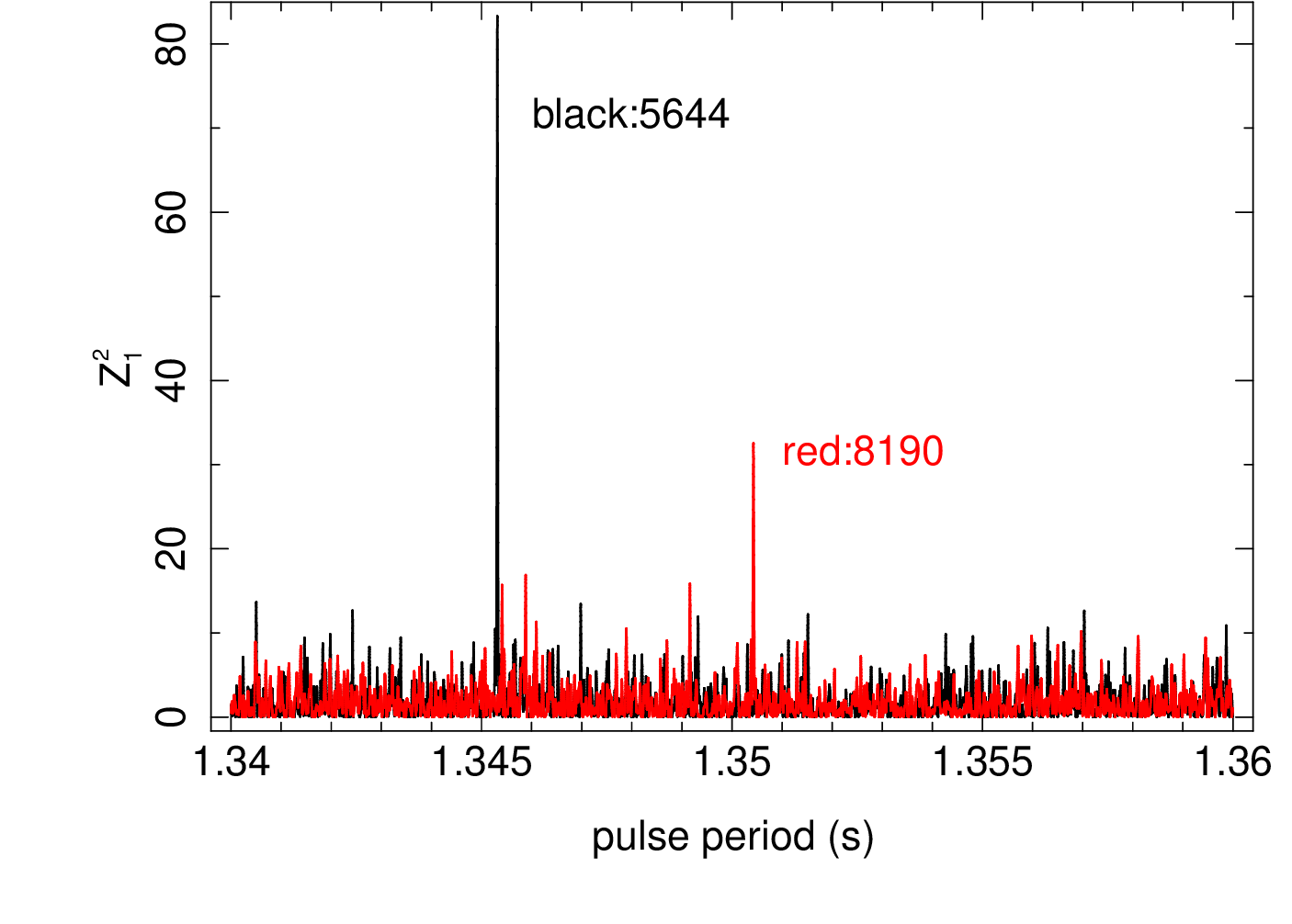}
	\includegraphics[width=3.2in]{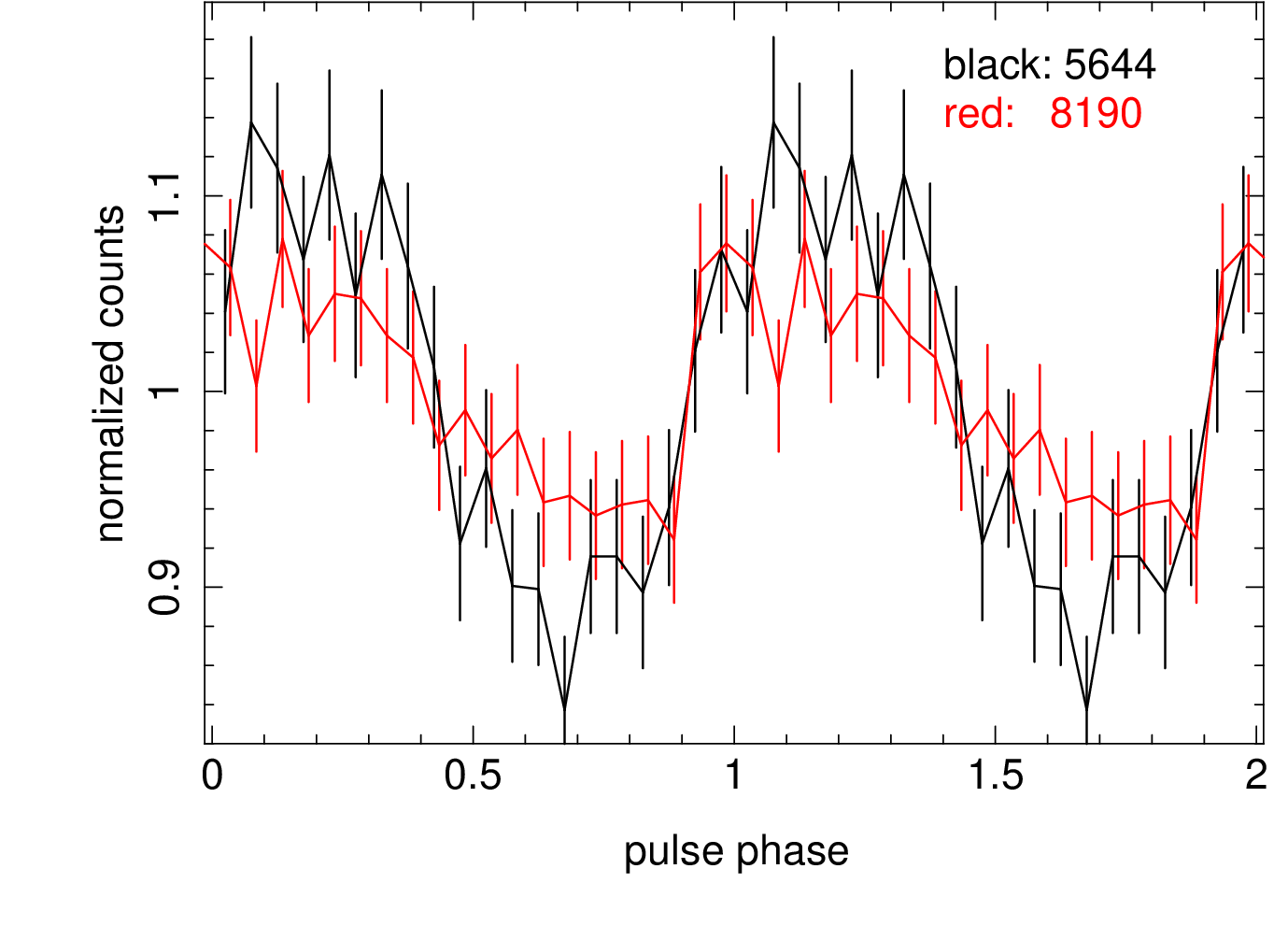}
	\includegraphics[width=3.2in]{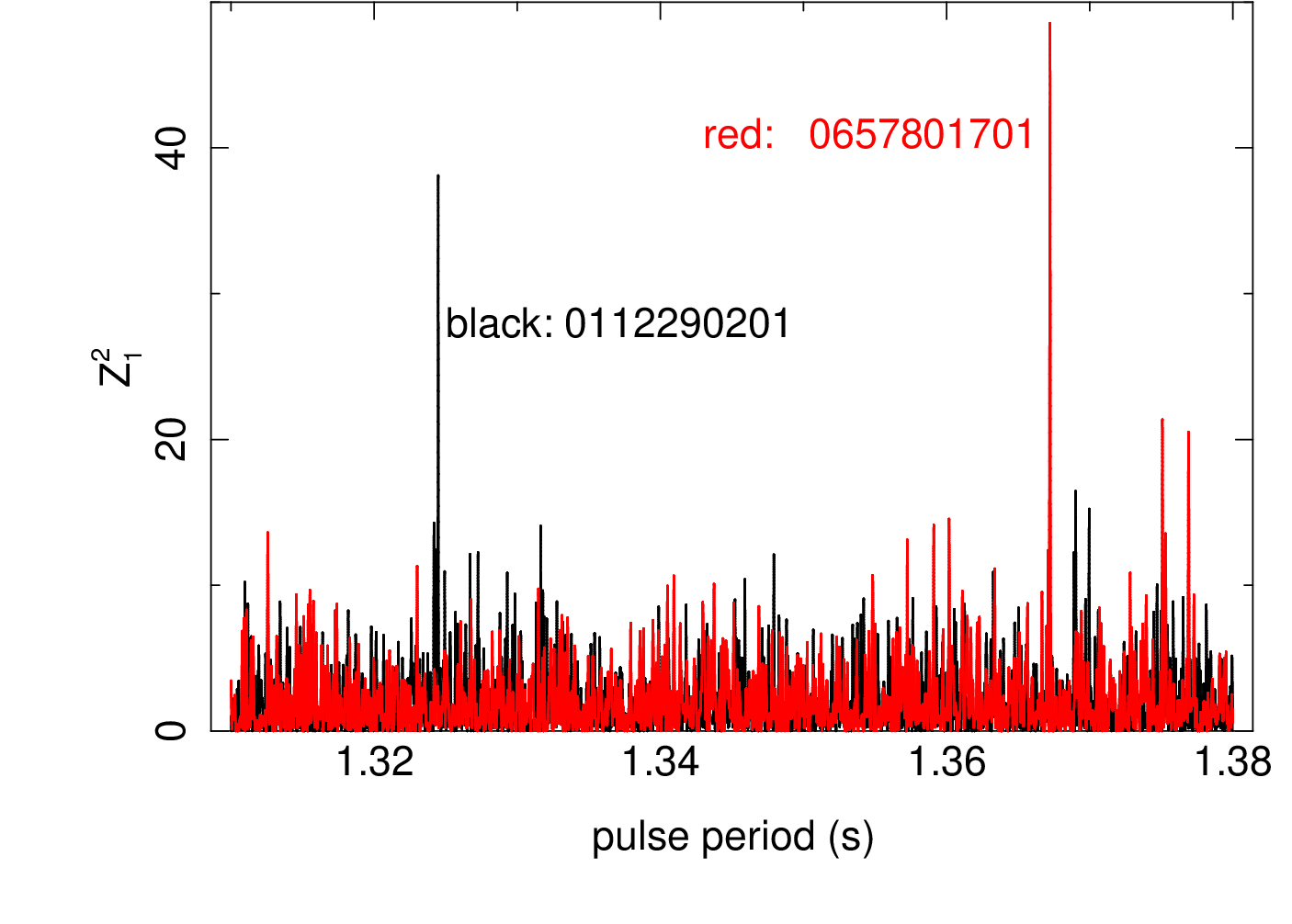}
	\hspace{1.5cm}
	\includegraphics[width=3.2in]{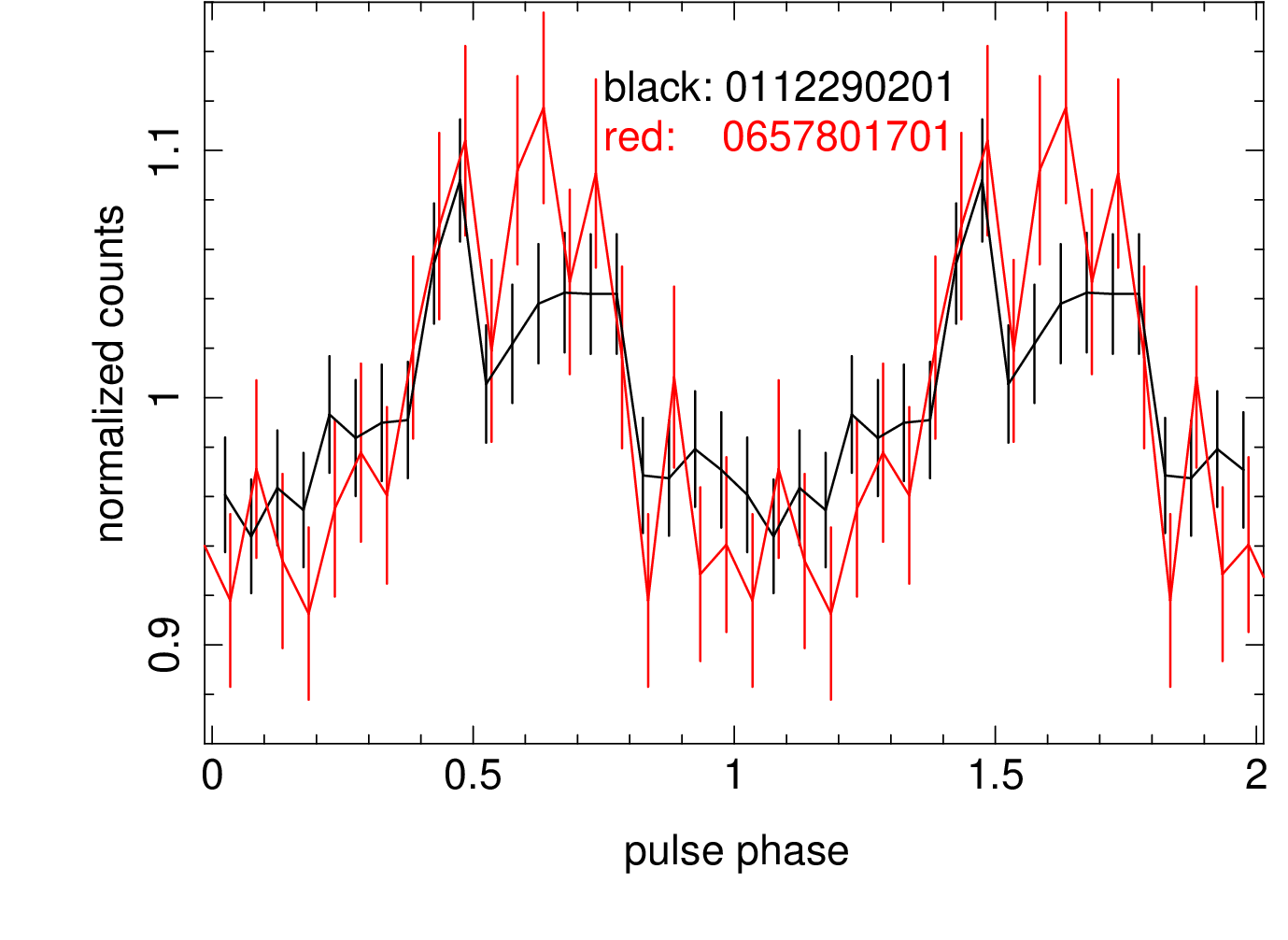}
	\caption{Top: pulsation detection in \cha dataset 5644 and 8190
within 2-8 keV band and their pulse profiles; bottom: 
pulsation detection in \xmm dataset 0112290201 and 0657801701
	within 3-8 keV band and their pulse profiles. For viewing purpose, the pulse profiles
	marked with red are right-shifted a little bit.
	} 
\end{figure*}

\section{Observational data}

The central region of M82 is very crowd and is generally dominated 
by M82 X-1 and X-2. Being separated by only $\sim5\arcsec$, 
currently only \cha can resolve M82 X-2 from M82 X-1.
The nominal full frame time of \cha ACIS is 3.2\,s, too high to detect 
a pulsation of period around 1.35\,s. The frame time of the \cha subarray mode is, however, 
smaller, with 0.44104\,s for ACIS-S. This makes 
\cha subarray data useful for the detection of pulsation from M82 X-2.
We searched all the \cha subarray observations of M82, using 
the standard data products created by the \cha team.
The events of M82 X-2 are extracted from a circle region of about $1.7\arcsec$
for on-axis observations, and from an ellipse for off-axis observations when the point
spread function (PSF) is elongated.
The extraction regions were chosen to avoid being blended with nearby sources.
The photon count rate is about 0.2\,s$^{-1}$, and we use the events data directly, without 
rebinning them into a light curve. The background signal is about 5\% of that of M82 X-2 and is not 
taken into account.

XMM-Newton EPIC-pn data have previously been shown to be useful for detecting pulsations from
M82 X-2 \citep{Bac22}. To supplement the \cha data, we also searched 
the \xmm EPIC-pn data before 2014. The \xmm data were reduced using the \xmm science analysis 
software (SAS, version 20). Circle regions of $30\arcsec$ were utilized, and we used 
the events data. We found two observational episodes showing significant 
pulsation, which are detailed in Table 1.
All our data were barycentered and corrected for the binary orbital effect 
using the orbital parameters obtained by \citet{Bac22}: 
$a{\rm sin}{\it i}=22.218$ ls, $T_{asc}=56682.06694$ (MJD), 
$P_{\rm orb}=2.5329733$ day,
$\dot{P}_{\rm orb}=-5.69\times10^{-8}$ s\,s$^{-1}$, and $e=0$.

\begin{table}
\scriptsize
   \begin{center}
\caption{\cha subarray observation and \xmm data}
\begin{tabular}{cccccc}
 \hline
	 Obs.ID & Date & MJD & T$_{exp}$ & Pulse period$^b$ & Sig$^c$\\
	  &  &  & (ks) &  (s) & ($\sigma$)\\
   \hline
	0112290201$^a$ & 2001-05-06 & 52035 & 31 & 1.324474(5) &4.0\\ 
	6097 &  2005-02-04 & 53406 & 53 & - &- \\
	5644 &  2005-08-17 & 53599 & 68 & 1.345317(2) &7.7\\
	6361 &  2005-08-18 & 53600 & 17 & - &-\\
	8190 &  2007-06-02 & 54253 & 53 & 1.350429(4) &3.3 \\
	10027 & 2008-10-04 & 54743 & 18 & - &-\\
	0657801701$^a$ & 2011-04-09 & 55660 & 24 & 1.367224(8)& 5.0 \\
 \hline
\end{tabular}
\begin{description}
  \begin{footnotesize}
  $^a$ obsIDs for \xmm data;\\
  $^b$ the quoted error is based on Monte-Carlo simulation of the time of arrival of photons 
	  from the pulse profile, a method similar as that used in \citet{Sin23};\\
  $^c$ the significance of detection in units of $\sigma$.
  \end{footnotesize}
   \end{description}
\end{center}
\end{table}

\section{Timing results}

We searched all \cha data of M82 with subarray configuration and found 5 observations 
for which M82 X-2 was bright, as listed in Table 1. 
We ran the $Z^2_n$ test \citep{Buc83} with $n=1$ to search for pulsation signals over all 
5 \cha ACIS subarray datasets. We used the 
Interactive Spectral Interpretation System \citep{ISIS}, and in particular,  
the $Z^2_n$ search routines provided in ISISscripts by the Remeis 
observatory\footnote{www.sternwarte.uni-erlangen.de/isis}.
We searched the period interval between 1.3 s and 1.4 s with a 
step size of 0.00001 s.
We also tried a longer interval and the result is unchanged.
We found that the step size needs to be refined to well sample the $Z^2_1$ peak.
The pulsed fraction of M82 X-2 is energy dependent \citep{Bac22}, and becomes smaller 
at lower energies. We tested different energy ranges for the pulsation signal and 
found that the 2-8 keV band generally provides the highest statistic. 
We therefore adopted the energy range of 2-8 keV for our analysis.

The dataset with the longest exposure, 5644,
has the highest test value $Z^2_1=83$ for the trial period of 1.345317(2) s, as shown 
in the top left panel of Figure 1. Such a high $Z^2_1$ value corresponds to a 
detection significance of about $7.7\sigma$. The uncertainty of the pulsation period 
is estimated by re-sampling the time of arrival of photons from the folded pulse profile.
The $Z^2_1$ value for dataset 8190 is 33 for a period of 1.350429(4) s, which is also shown in Figure 1.
The pulse profiles of both datasets folded with the periods determined are shown 
in the top right panel of Figure 1, and they 
are very similar to those reported in \citet{Bac14}.
To estimate the pulse fraction of a profile, we fitted it with a sinusoidal function with 
only the fundamental component. The pulse fractions (defined as the semi-amplitude of the sinusoid 
divided by the mean count rate) are $12\pm2$\% and $6\pm2$\%, for dataset 5644 and 8190, respectively.
We note that the background is negligible for the on-axis observation 5644, but for the off-axis 
observation 8190, M82 X-2 is blended with nearby sources (not M82 X-1) and its pulse fraction
is only a lower limit. 
We found no maximum statistics of the other three datasets above 30, and they
are neglected hereafter.

For \xmm data, the events of M82 X-2 are contaminated with those of M82 X-1, and we found that the 
energy range of 3-8 keV 
is more sensitive to the pulsation signal. Searching for the 3-8 keV events, we 
found two significant $Z^2_1$ peaks, 
for observations 0112290201 and 0657801701.
The resulting $Z^2_1$ statistics and pulse profiles are plotted in 
the bottom panels of Figure 1.
The pulse profiles are similar to those of \cha data, with a 
pulse fraction of $5\pm1$\% and $8\pm2$\%, for 0112290201 and 0657801701, respectively. 
We note that the background level within 3-8 keV of \xmm data is around 7\%, and 
the fluxes of M82 X-2 of \xmm data include 
contributions from other sources (not only M82 X-1). These contaminations are hard 
to estimate without simultaneous 
\cha observation and may be as luminous as M82 X-2 itself. Therefore, the estimated pulse fraction 
from \xmm data is only a lower limit.

Our newly measured spin frequencies are plotted in Figure 2, together
with those measured from \nus data by \citet{Bac22}. As can be seen, 
over the 20 years timescale, the spin frequency of M82 X-2 is decreasing 
continuously. The spin-down rate seems reduced a little around 2021 (MJD 59000),
and then returned to its more normal value.	
The four new detections approximately follow
a spin-down rate of $-7.4\times10^{-11}$ Hz\,s$^{-1}$, as shown by the 
dotted line in Figure 2. We note that the four new detections can not be well fitted with a 
linear function, as the resulting residuals are too large.

\begin{figure}
	\hspace{-0.2in}
	\includegraphics[width=3.5in]{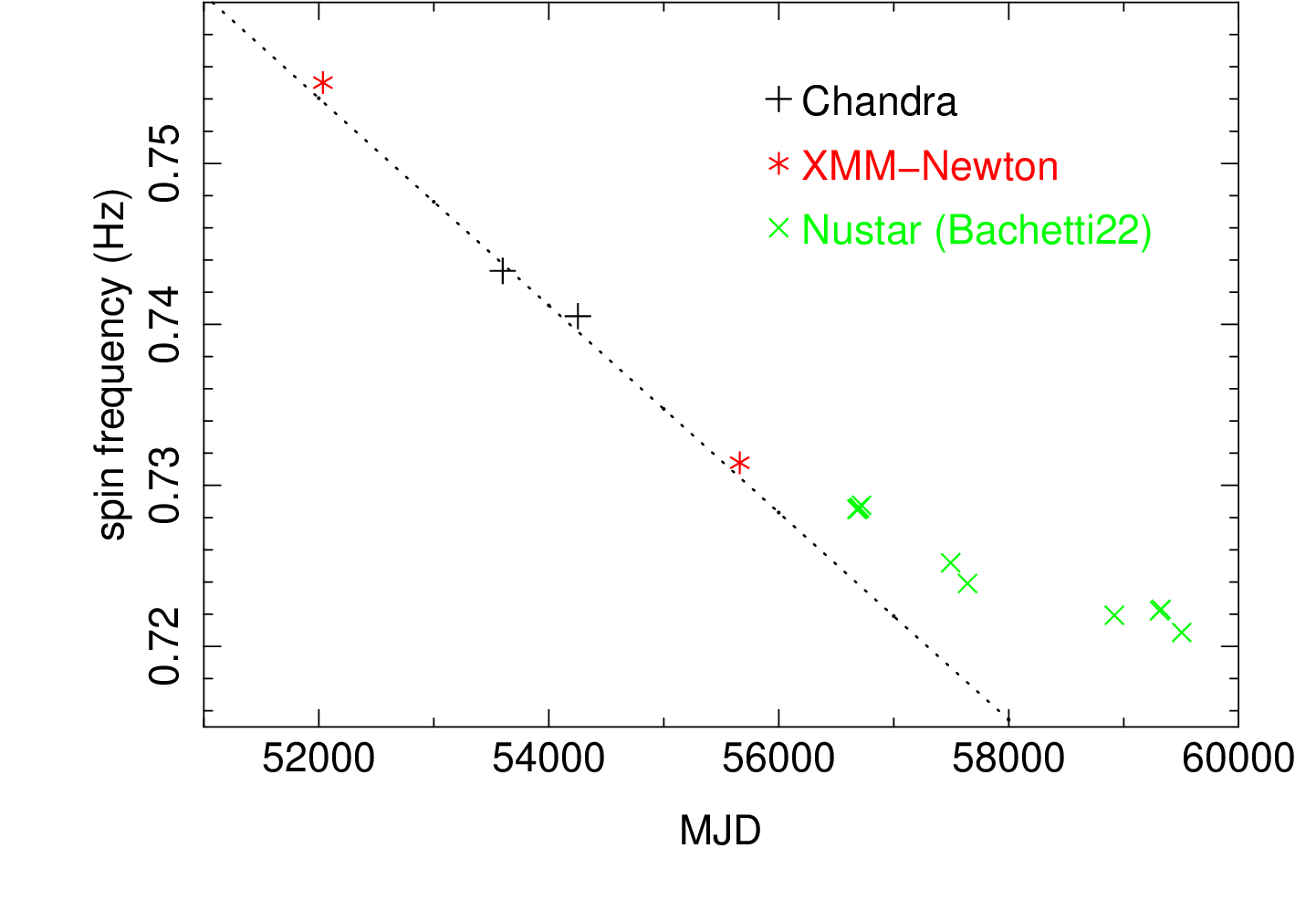}
	\caption{
Spin history of M82 X-2 since 2001. The newly measured frequencies clearly indicate a long-term 
spin-down trend, with a spin-down rate of $-7.4\times10^{-11}$Hz\,s$^{-1}$, which is 
shown as the dotted line. The uncertainties of the measurements are too small to be illustrated.
}
\end{figure}


The exposure of \cha dataset 5644 is about 70\,ks, which is long enough to allow us to 
constrain the spin frequency derivative ($\dot{\nu}$).
We divided the exposure into 2 intervals of 35 ks and measured their 
pulsations separately.
We obtained pulse periods of 
1.345321(5) s and 1.345306(4) s around MJD 53599.28 and 53599.71, respectively.
These correspond to an averaged spin-up rate of $\sim2.2\times10^{-10}$ Hz\,s$^{-1}$.


\section{Discussion and conclusion}

We measured the pulsation of M82 X-2 back to 2005 and 2001 using 
\cha and \xmm archive data.
With our newly determined spin frequencies, M82 X-2 shows a clear spin-down trend
over a timescale of 20 years, with occasional spin-up events. 
Such spin behavior is similar to that of Be-type X-ray binaries (BeXBs).
BeXBs generally show a spin-up trend during an outburst or giant outburst state, and 
show a continuous spin-down trend when there is much less mass to be accreted.
A typical example is the first Galactic ULXP, Swift J0243.4+6126, which was 
discovered during a giant outburst in Oct. 2017 \citep[e.g.][]{Wil18}.
After the giant outburst, Swift J0243 showed a continuous spin-down trend
(with a rate $\sim-2\times10^{-12}$Hz/s)
with its flux being low \citep{Liu23}.

It is known that M82 X-2 flux shows 
strong variations and may have a varying period around 60 days \citep[e.g.][]{Bri19}. 
\citet{Bri19} found that the overall spectral shape of M82 X-2 is quite similar at 
low and high levels and the flux variations are unlikely caused by occultations.
The changing spin-down/spin-up behavior of M82 X-2 indicates that its accretion torque, 
and thus its accretion rate, must be varying. 
Therefore, it is possible that  
M82 X-2 may have a varying accretion rate, and for long periods of time it is in a relatively quiet state
and spins down. 
A spin-down trend is expected when the accretion rate is low enough that the inner 
disk radius is larger than the co-rotation radius ($R_c$), and the magnetic
field threading produces a braking torque.

As the ellipticity of M82 X-2 is very small, the origin of the accretion variation 
of M82 X-2 should be different from the regular type I outbursts from BeXBs.
One potential origin of variation is the optical donor star. 
As recently reported, some classical X-ray pulsars show 
alternating spin-up/spin-down torque reversals on tens of days \citep{Liao22a,Liao22b}.
It was found that for OAO 1657-415 and Vela X-1 their orbital profiles 
and accretion flows on the orbital scale are different for different torque states,
indicating a variation of the flow from the optical donor star over tens 
of days. Similarly, the accretion flow from M82 X-2's optical star may also vary.

If the long-term spin-down trend of M82 X-2 is caused by magnetic threading, 
we can infer a dipolar magnetic field based on the magnetic braking torque, 
as we did for Swift J0243 \citep{Liu23}. Taking the braking torque as 
$\tau_{b}=-\mu^2/9R_c^3$ \citep{Wang95,Rap04}
and neglecting the mass accretion torque,
we can infer a dipolar magnetic field of $B\sim1.2\times10^{13}$ G,
using $2\pi I\dot{\nu}=\tau_{b}$ and assuming a neutron star
of 1.4 $M_\odot$ with a radius of 10 km. This dipolar field is similar
to that of Swift J0243 ($B\sim1.75\times10^{13}$ G).
While the spin-down rate of M82 X-2 ($-7.4\times10^{-11}$Hz/s) is 
35 times higher than that of Swift J0243, its $R_c$ is 3.7 times smaller 
than that of Swift J0243, which makes the braking torque more efficient.

With the estimated dipolar field of M82 X-2 above, we can compare 
the braking torque with the accretion torque. During the observation of 5644, 
the average spin-up rate is about $2.2\times10^{-10}$Hz/s, which is about 
twice as high as the average value during the 2014 observations \citep{Bac14}.
The unabsorbed 0.5-10 keV luminosity of 5644 is about $1\times10^{40}$erg/s
\citep{Bri16}. At such a high luminosity, the accretion disk should be in a 
radiation pressure dominated state, and the inner disk radius should change
very slowly with the accretion rate \citep{Cha17,Cha19}:
\begin{equation}
	R_{in}\simeq3.6\times10^7\mu_{30}^{4/9}\sim8\times10^7 \rm cm,
\end{equation}
where $\mu_{30}$ is the magnetic moment in units of $10^{30}$G\,cm$^3$.
To produce the observed spin-up rate, a mass accretion 
rate of $\sim1.3\times10^{19}$ g\,s$^{-1}$
($2\pi I\dot{\nu}=\dot{m}\sqrt{GMR_{in}}$) is required, which is about 
four times less than that inferred from the observed luminosity.
That is, a mild beaming factor of about four is needed to match the accretion torque
with the braking torque.

However, the long-term spin-down trend of M82 X-2 may be not caused 
by magnetic threading. Many Galactic X-ray pulsars 
show alternating spin-up/spin-down reversals, such as OAO 1657-415 \citep{Liao22a}.
The flux of OAO 1657-415 is correlated with the spin-up rate and 
anti-correlated with the spin-down rate, implying that its torque reversals 
could be caused by alternating prograde/retrograde
accretion flows to the neutron star. A retrograde flow will lead to a spin-down 
whereas a prograde flow will lead to a spin-up. This implies that
the accretion flow (mass transfer) to the neutron star is unstable or variable.
One possible scenario is 
the irradiation-driven instability of a processing warped disk, which may lead to 
a flip-over inner disk, as first proposed for Cen X-3 by \citet{van98}.

It is interesting to note that \citet{Bac22} reported some \nus
observations of instantaneous spin-down behaviors. As \nus cannot resolve M82 X-2 from X-1, 
to make its pulsation detectable by \nus data,
the fluxes of M82 X-2 would need to be relatively high. In the magnetic braking scenario,
to be in a spin-down state, the accretion rate/fluxes should be relatively low.
While there were no exact simultaneous \cha data 
for the \nus observations of spin-down, we found that 
for one \nus observation (obsID 30702012002, on MJD 59505, 
the last data point in Figure 2), the \cha data
observed two days later (obsID 23471) do show a flux ratio between M82 X-2 
and X-1 similar to that of data 5644.
If confirmed, such spin-down observations of high M82 X-2 fluxes will 
support a non-magnetic origin for the spin-down of M82 X-2, and 
a retrograde flow to the neutron star would be preferred.
Further monitoring of M82 X-2 with higher cadence using the \cha subarray mode, 
including both high and low states, is required to reveal 
the true long-term spin-down nature of M82 X-2.

If the spin-down trend always held in the past, one can infer a spin-down 
time of $\nu/\dot{\nu}\sim10^5$ year. Such a time is consistent with 
the young, star-bursting nature of M82, and may 
correspond to the formation time of the neutron star of M82 X-2. 
Nevertheless, we note that
spin-up/spin-down reversals on a timescale of years have been observed in X-ray pulsars 
\citep[such as LMC X-4][]{Mol17}, and one should be cautious when inferring a longer-term trend.


\section*{Acknowledgements}
We thank the referee for his/her thoughtful comments which have improved the paper considerably
and Richard Long for a through reading.
This work used data from \xmm telescope and
employed a list of Chandra datasets, obtained by the Chandra X-ray Observatory, 
contained in~\dataset[DOI: cdc.172]{https://doi.org/10.25574/cdc.172}.
This research has made use of a collection of ISIS functions (ISISscripts) provided 
by ECAP/Remeis observatory and MIT (http://www.sternwarte.uni-erlangen.de/isis/).
We acknowledge the support by National 
Natural Science Foundation of China (U1938113),
the Scholar Program of Beijing Academy of Science and Technology (DZ BS202002),
and the science research grants from the China Manned Space Project.


%
\bibliographystyle{mn2e}

\end{document}